\documentclass[preprint,showpacs,preprintnumbers,amsmath,amssymb]{revtex4}
\usepackage{booktabs}
\usepackage{mathrsfs}
\usepackage{epsfig}
\usepackage{graphicx}
\usepackage{dcolumn}
\usepackage{bm}
\usepackage{amsmath}

\let\jnfont=\rm
\def\NPB#1,{{\jnfont Nucl.\ Phys.\ B }{\bf #1},}
\def\PLB#1,{{\jnfont Phys.\ Lett.\ B }{\bf #1},}
\def\EPJC#1,{{\jnfont Eur.\ Phys.\ Jour.\ C }{\bf #1},}
\def\PRD#1,{{\jnfont Phys.\ Rev.\ D }{\bf #1},}
\def\PRL#1,{{\jnfont Phys.\ Rev.\ Lett.\ }{\bf #1},}
\def\MPLA#1,{{\jnfont Mod.\ Phys.\ Lett.\ A }{\bf #1},}
\def\JPG#1,{{\jnfont J.\ Phys.\ G}{\bf #1},}
\def\CTP#1,{{\jnfont Commun.\ Theor.\ Phys.\ }{\bf #1},}
\def\ZPC#1,{{\jnfont Z.\ Phys.\ C }{\bf #1},}
\def\JHEP#1,{{\jnfont JHEP \ }{\bf #1},}
\def\Rv{\not{\hbox{\kern-1pt $R$}}}
\def\p{\not{\hbox{\kern-3pt $p$}}}

\newcommand{\SLASH}[2]{\makebox[#2ex][l]{$#1$}/}
\newcommand{\pslash}{\SLASH{p}{.2}}

\begin{document}
\preprint{\parbox{1.2in}{\noindent arXive:0908.4556}}

\title{Lepton flavor-changing processes in $R$-parity violating MSSM:
       $Z\to \ell_i\bar\ell_j$ and $\gamma\gamma\to \ell_i\bar\ell_j$
       under new bounds from $\ell_i\to \ell_j \gamma$}

\author{Junjie Cao$^1$, Lei Wu$^1$, Jin Min Yang$^{2,3}$ \\~ \vspace*{-0.5cm} }
\affiliation{
$^1$ College of Physics and Information Engineering, Henan Normal University,
     Xinxiang 453007, China\\
$^2$ Key Laboratory of Frontiers in Theoretical Physics, Institute of Theoretical Physics,
     Academia Sinica, Beijing 100190, China \\
$^3$ Kavli Institute for Theoretical Physics China, Academia Sinica, Beijing 100190, China
     \vspace*{1.5cm}}

\begin{abstract}
We examine the lepton flavor-changing processes in $R$-parity violating MSSM.
First, we update the constraints on the relevant $R$-violating couplings
by using the latest data on the rare decays $\ell_i\to \ell_j \gamma$.
We find that the updated  constraints are much stronger than the old ones from rare
$Z$-decays at LEP.
Then we calculate the processes $Z \to \ell_i \overline{\ell}_j$ and
$\gamma\gamma\to \ell_i\bar\ell_j$. We find that with the updated  constraints
the  $R$-violating couplings can still enhance the rates of these processes
to the sensitivity of GigaZ and photon-photon collision options of the ILC.
\end{abstract}


\maketitle

\section{INTRODUCTION}
The minimal supersymmetric model (MSSM) is a popular extension of
the Standard Model (SM). In this model the invariance of $R$-parity,
defined by $R=(-1)^{2S+3B+L}$ for a field with spin $S$,
baryon-number $B$ and  lepton-number $L$,  is often imposed on the
Lagrangian in order to maintain the separate conservation of
baryon-number and lepton-number. Although $R$-parity plays a crucial
role in the phenomenology of the MSSM (e.g., forbid proton decay and
ensure a perfect candidate for cosmic dark matter), it is, however,
not dictated by any fundamental principle such as gauge invariance
and there is no compelling theoretical motivation for it. The most
general superpotential of the MSSM consistent with the SM gauge
symmetry and supersymmetry contains $R$-violating interactions which
are given by~\cite{rp1}
\begin{equation}\label{poten}
{\cal W}_{\not \! R}=\frac{1}{2}\lambda_{ijk}L_iL_jE_k^c
+\lambda'_{ijk} L_iQ_jD_k^c
+\frac{1}{2}\lambda''_{ijk}\epsilon^{abd}U_{ia}^cD_{jb}^cD_{kd}^c
+\mu_iL_iH_2,
\end{equation}
where $i,j,k$ are generation indices, $c$ denotes charge
conjugation, $a$, $b$ and $d$ are the color indices with
$\epsilon^{abd}$ being the total antisymmetric tensor,  $H_{2}$ is
the Higgs-doublet chiral superfield, and $L_i(Q_i)$ and
$E_i(U_i,D_i)$ are the left-handed lepton (quark) doublet and
right-handed lepton (quark) singlet chiral superfields. The
dimensionless coefficients $\lambda_{ijk}$
(antisymmetric in $i$ and $j$) and $\lambda'_{ijk}$
 in the superpotential are
$L$-violating couplings, while $\lambda''_{ijk}$ (antisymmetric in
$j$ and $k$) are $B$-violating couplings. So far both theorists and
experimentalists have intensively studied the phenomenology of
$R$-parity breaking supersymmetry in various processes
\cite{rp2,rp3} and obtained some bounds \cite{review}.

The lepton flavor-changing (LFC) processes, which have been
searched in various experiments \cite{exp2,exp3,exp4}, are a
sensitive probe for new physics because they are extremely
suppressed in the SM but can be greatly enhanced in new physics
models like supersymmetry \cite{LFC-1}. In $R$-parity breaking
supersymmetry, these rare processes may receive exceedingly large
enhancement since both $\lambda$ and $\lambda^{\prime}$ couplings
can make contributions. Such enhancement was considered in the
decays $l_i \to l_j \gamma$ \cite{liljgamma} and $Z \to \ell_i
\bar\ell_j$ \cite{Z-decay}, the $\mu-e$ conversion in
nuclei \cite{mueconversion}, and the di-lepton productions $p
\bar{p}/pp \to e^\pm \mu^\mp +X $ \cite{l-prod-hadron}
and  $e^+ e^- \to e^\pm \mu^\mp$ \cite{l-prod-ILC}.

Since the GigaZ and photon-photon collision options of the ILC can
precisely measure the LFC processes $Z \to \ell_i \bar\ell_j$ and
$\gamma \gamma \to \ell_i \bar\ell_j$ ($i\neq j$ and
$\ell_i=e,\mu,\tau$), we in this work study these processes in
$R$-violating MSSM. Noting that the experimental upper bounds on the LFC
$\tau$-decays became more stringent recently \cite{exp3}, we
will first check the constraints on the relevant $R$-violating
couplings from the latest measurement of $\ell_i \to \ell_j \gamma$.
Then, with the updated bounds on the relevant $R$-violating
couplings, we calculate $Z \to \ell_i \bar\ell_j$ and $\gamma \gamma
\to \ell_i \bar\ell_j$ to figure out if they can reach the
sensitivity of the GigaZ and photon-photon collision options of the
ILC.

The paper is organized as follows. In Sec. II we describe the calculations for
$\ell_i \to \ell_j \gamma$, $Z \to \ell_i \bar\ell_j$ and
$\gamma \gamma \to \ell_i \bar\ell_j$.
In Sec. III we present some numerical results and discussions.
Finally, a conclusion is drawn in Sec. IV.

\section{Calculations}
In terms of the four-component Dirac notation, the Lagrangian of the $L$-violating
interaction is given by (in our calculations we take the presence of $\lambda_{ijk}'$
as an example)
\begin{eqnarray}
{\cal L}_{\lambda^{\prime}}&=&-\lambda^{\prime}_{ijk}
  \left [\tilde \nu^i_L \overline{d^k_R} d^j_L
  + \tilde d^j_L  \overline{d^k_R} \nu^i_L
  + (\tilde d^{k}_R)^*  \overline{(\nu^i_L)^c} d^j_L \right. \nonumber\\
&& ~~~~~~~~
  \left.- \tilde l^i_L  \overline{d^k_R} u^j_L
  - \tilde u^j_L  \overline{d^k_R} l^i_L
  - (\tilde d^k_R)^* \overline{(l^i_L)^c} u^j_L\right]+h.c.
\end{eqnarray}
The LFC interactions $\ell_i \bar\ell_j V$ ($V=\gamma,Z$) are
induced at loop level by exchanging a squark $\tilde u^j_L$ or
$\tilde d^{k}_R$, which is shown in Fig.1.

\begin{figure}[htbp]
\epsfig{file=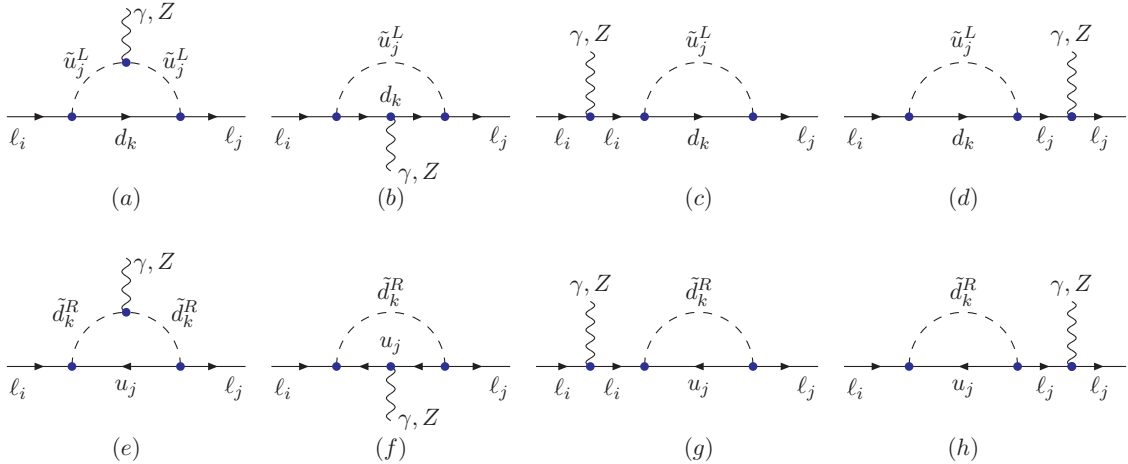,width=15cm} \vspace*{-0.7cm}
\caption{Feynman diagrams for $\ell_i - \ell_j $ transition induced by
        the $L$-violating couplings at one-loop level.}
\label{fig1}
\end{figure}
\begin{figure}[htbp]
\epsfig{file=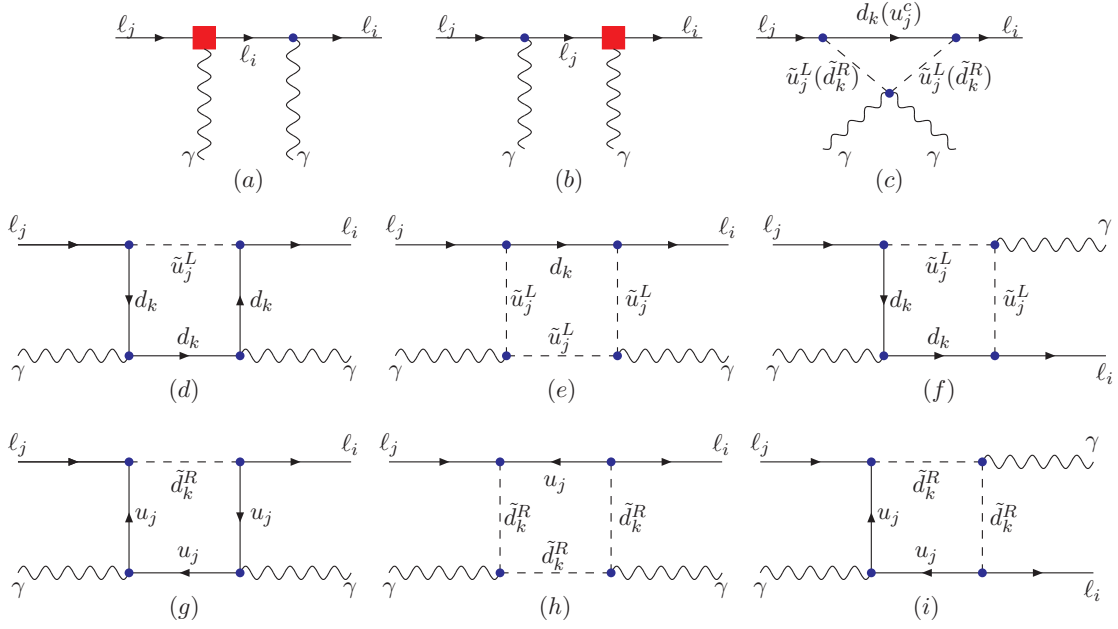,width=15cm} \vspace*{-0.7cm} \caption{Feynman
diagrams for $\gamma \gamma \to \ell_i \bar\ell_j$ induced by
the $L$-violating couplings at one-loop level.
The effective $\gamma-\ell_i-\ell_j$ vertex in (a,b) is defined in Fig. 1.}
\label{fig2}
\end{figure}

For the decays $\ell_i \to \ell_j \gamma$ we take $\mu \to e \gamma$
as an example to show the analytic results.
The gauge invariant amplitude of $\mu \to e \gamma$ is given by
\begin{eqnarray}
 M(\mu \to e \gamma)&=& 2 A \bar{u}(p_{e})P_R(2\epsilon \cdotp p_\mu
      - m_\mu \cdotp \epsilon\!\!\slash)u(p_{\mu}) ,
\end{eqnarray}
where $A$ is given by (assuming the degeneracy for squark masses)
\begin{eqnarray}
A&=&\frac{ie\lambda^{\prime}_{1jk}\lambda^{\prime}_{2jk}}{16\pi^{2}}
  \frac{m_\mu}{ m^2_{\tilde{q}}} \left[
  f_1(\frac{m_{d_k}^2}{m^2_{\tilde{q}}}) +
  f_2(\frac{m_{u_j}^2}{m^2_{\tilde{q}}}) \right]
\end{eqnarray}
with
\begin{eqnarray}
f_1(x)&=&\frac{1}{8(x -1)^3}\left[
  \frac{2}{3}(2 x^2 +5 x -1-\frac{6x^2 \ln x}{x -1})
  -\frac{1}{3}(x^2 -5x -2+\frac{6x  \ln x}{x -1})\right] ,\\
f_2(x)&=& \frac{1}{8(x-1)^3}\left[\frac{1}{3}(2x^2+5x-1-\frac{6x^2
\ln x}{x-1})
  -\frac{2}{3}(x^2-5x-2+\frac{6x \ln x}{x-1})\right] .
\end{eqnarray}
The decay branching ratio reads
\begin{eqnarray}
BR(\mu \to e \gamma)=\frac{48\pi}{G_F^2m_\mu^2}|A|^2 .
\end{eqnarray}
For the decays $Z\to \ell_i \bar\ell_j$ we calculate the decay
rates numerically by using the effective vertex presented in
Appendix A. Note that according to the effective vertex method
\cite{Cao}, the external legs of the effective vertex can be
on-shell or off-shell and thus the vertex can be used in any
relevant process. The expression in Eqs.(3-6) can be obtained from
the effective vertex in Appendix A by putting both leptons on
shell.

For the process $\gamma\gamma \to \ell_i \bar\ell_j$, besides
Fig.2 (a,b) induced by the effective vertex given in Appendix A,
more diagrams shown in Fig.2 (c-i) also come into play.
The analytic expressions of the amplitudes of these diagrams
are given in Appendix B. These amplitudes contain the Passarino-Veltman
one-loop functions, which are calculated by using LoopTools \cite{Hahn}.
We checked that the amplitudes have gauge invariance and
the ultraviolet divergence cancelled.

Since the photon beams in $\gamma\gamma$ collision are generated
by the backward Compton scattering of the incident electron- and
the laser-beam, the events number is obtained by convoluting the
cross section of $\gamma\gamma$ collision with the photon beam
luminosity distribution:
\begin{eqnarray}
N_{\gamma \gamma \to \ell_i \bar\ell_j}&=&\int d\sqrt{s_{\gamma\gamma}}
  \frac{d\cal L_{\gamma\gamma}}{d\sqrt{s_{\gamma\gamma}}}
  \hat{\sigma}_{\gamma \gamma \to \ell_i \bar\ell_j}(s_{\gamma\gamma})
  \equiv{\cal L}_{e^{+}e^{-}}\sigma_{\gamma \gamma \to \ell_{i} \bar\ell_{j}}(s)
\end{eqnarray}
where $d{\cal L}_{\gamma\gamma}$/$d\sqrt{s}_{\gamma\gamma}$ is the photon-beam luminosity
distribution and $\sigma_{\gamma \gamma \to \ell_i \bar\ell_j}(s)$ (
$s$ is the squared center-of-mass energy of $e^{+}e^{-}$ collision) is defined as
the effective cross section of $\gamma \gamma \to \ell_{i} \bar\ell_{j}$.
In optimum case, it can be written as \cite{photon collider}
\begin{eqnarray}
\sigma_{\gamma \gamma \to \ell_i \bar\ell_j}(s)&=&
  \int_{\sqrt{a}}^{x_{max}}2zdz\hat{\sigma}_{\gamma \gamma \to \ell_{i} \bar\ell_{j}}
  (s_{\gamma\gamma}=z^2s) \int_{z^{2/x_{max}}}^{x_{max}}\frac{dx}{x}
 F_{\gamma/e}(x)F_{\gamma/e}(\frac{z^{2}}{x})
\end{eqnarray}
where $F_{\gamma/e}$ denotes the energy spectrum of the back-scattered photon for the
unpolarized initial electron and laser photon beams given by
\begin{eqnarray}
F_{\gamma/e}(x)&=&\frac{1}{D(\xi)}\left[1-x+\frac{1}{1-x}-\frac{4x}{\xi(1-x)}
  +\frac{4x^{2}}{\xi^{2}(1-x)^{2}}\right]
\end{eqnarray}
with
\begin{eqnarray}
D(\xi)&=&(1-\frac{4}{\xi}-\frac{8}{\xi^{2}})\ln(1+\xi)
  +\frac{1}{2}+\frac{8}{\xi}-\frac{1}{2(1+\xi)^{2}}.
\end{eqnarray}
Here $\xi=4E_{e}E_{0}/m_{e}^{2}$ ($E_{e}$ is the incident
electron energy and $E_{0}$ is the initial laser photon energy)
and $x=E/E_{0}$ with $E$ being the energy of the scattered photon
moving along the initial electron direction.

\section{Numerical results and discussions }
In our calculations we take the SM parameters as \cite{pdg}
\begin{eqnarray}
m_{\mu}=0.106{\rm ~GeV}, m_{\tau}=1.777{\rm ~GeV}, m_{b}=4.2{\rm ~GeV}, \alpha=1/137,
\sin^2\theta_W=0.223
\end{eqnarray}
The top quark mass is taken as the new CDF value $m_{t}=172.3{\rm ~GeV}$ \cite{CDF-mt}.
The relevant SUSY parameters in our calculations are the masses of squarks
as well as the $R$-parity violating couplings listed in Table I.
The strongest bound on squark mass is from the Tevatron experiment.
For example, from the search of the inclusive production of squark and gluino
in $R$-conserving minimal supergravity model with $A_0 = 0$, $\mu< 0$ and $\tan \beta = 5$,
the CDF gives a bound of 392 GeV at the 95 $\%$ C.L. for degenerate gluinos and
squarks \cite{CDF}. However, this bound may be not applicable to the $R$-violating
scenario because the SUSY signal in case of $R$-violation is very different from
the $R$-conserving case. The most robust bounds on sparticle masses come from
the LEP results, which give a bound of about 100 GeV on squark or slepton mass \cite{LEP}.
In our numerical calculations, we assume the presence of the minimal number of $R$-violating
couplings, i.e., for each process only the two relevant couplings
(not summed over the family indices) are assumed to be present.

For $\ell_i \to \ell_j \gamma$, the latest experimental data is \cite{exp4}
\begin{eqnarray}
BR(\mu \to e \gamma)&<&1.2\times 10^{-11},\\
BR(\tau \to e \gamma)&<&1.1\times 10^{-7},\\
BR(\tau \to \mu \gamma)&<& 4.5\times 10^{-8}.
\end{eqnarray}
We use these data to update the bounds on the relevant $L$-violating couplings.
The new bounds are compared with the old ones in Table I for
$m_{\tilde{q}}=100$ GeV (here we take squark mass of 100 GeV for illustration
and for heavier squarks the bounds on the $L$-violating couplings will become weak,
as will be shown later).
We can see that the new bounds are much stronger than the old ones.
Since the bounds on $\lambda_{i33}^{'}\lambda_{j33}^{'}$ ($i\neq j$) are weakest,
we only consider the contribution of $\lambda_{i33}^{'}\lambda_{j33}^{'}$ in our
following numerical calculations.

\begin{table}
\caption{Our new upper bounds on the $L$-violating couplings for $m_{\tilde{q}}=100$ GeV
         from $\ell_i \to \ell_j \gamma$ data \cite{exp4}, in comparison with the old ones \cite{review}.}
 \begin{tabular}{lll} \hline
 couplings &~~~~~~~~~~~~~New bounds& ~~~~~~~~~~Old bounds \cite{review}\\
 \hline
$\lambda^{\prime}_{111}\lambda^{\prime}_{211}$,~~$\lambda^{\prime}_{112}\lambda^{\prime}_{212}$ &~~~~~~~~~~~~~~ $7.74\times 10^{-5}$ &
~~~~~~~~~~~~~~~~$5.7\times 10^{-4}$ \\
$\lambda^{\prime}_{113}\lambda^{\prime}_{213}$ &~~~~~~~~~~~~~~ $7.85\times 10^{-5}$ &
~~~~~~~~~~~~~~~~$5.7\times 10^{-4}$ \\
$\lambda^{\prime}_{121}\lambda^{\prime}_{221}$,~~$\lambda^{\prime}_{122}\lambda^{\prime}_{222}$ &~~~~~~~~~~~~~~ $7.78\times 10^{-5}$ &
~~~~~~~~~~~~~~~~$5.7\times 10^{-4}$ \\
$\lambda^{\prime}_{123}\lambda^{\prime}_{223}$ &~~~~~~~~~~~~~~ $7.89\times 10^{-5}$ &
~~~~~~~~~~~~~~~~$5.7\times 10^{-4}$ \\
$\lambda^{\prime}_{131}\lambda^{\prime}_{231}$,~~$\lambda^{\prime}_{132}\lambda^{\prime}_{232}$ &~~~~~~~~~~~~~~ $1.27\times 10^{-3}$ &
~~~~~~~~~~~~~~~~$7.7\times 10^{-3}$ \\
$\lambda^{\prime}_{133}\lambda^{\prime}_{233}$ &~~~~~~~~~~~~~~ $1.63\times 10^{-3}$ &
~~~~~~~~~~~~~~~~$1.0\times 10^{-2}$ \\
$\lambda^{\prime}_{111}\lambda^{\prime}_{311}$,~~$\lambda^{\prime}_{112}\lambda^{\prime}_{312}$ &~~~~~~~~~~~~~~ $5.54\times 10^{-4}$ &
~~~~~~~~~~~~~~~~$1.2\times 10^{-2}$ \\
$\lambda^{\prime}_{113}\lambda^{\prime}_{313}$ &~~~~~~~~~~~~~~ $5.56\times 10^{-4}$ &
~~~~~~~~~~~~~~~~$1.2\times 10^{-2}$ \\
$\lambda^{\prime}_{121}\lambda^{\prime}_{321}$,~~$\lambda^{\prime}_{122}\lambda^{\prime}_{322}$ &~~~~~~~~~~~~~~ $5.57\times 10^{-4}$ &
~~~~~~~~~~~~~~~~$1.2\times 10^{-2}$ \\
$\lambda^{\prime}_{123}\lambda^{\prime}_{323}$ &~~~~~~~~~~~~~~ $5.65\times 10^{-4}$ &
~~~~~~~~~~~~~~~~$1.2\times 10^{-2}$ \\
$\lambda^{\prime}_{131}\lambda^{\prime}_{331}$,~~$\lambda^{\prime}_{132}\lambda^{\prime}_{332}$ &~~~~~~~~~~~~~~ $9.06\times 10^{-3}$ &
~~~~~~~~~~~~~~~~$1.2\times 10^{-2}$ \\
$\lambda^{\prime}_{133}\lambda^{\prime}_{333}$ &~~~~~~~~~~~~~~ $1.17\times 10^{-2}$ &
~~~~~~~~~~~~~~~~$1.2\times 10^{-2}$ \\
$\lambda^{\prime}_{211}\lambda^{\prime}_{311}$,~~$\lambda^{\prime}_{212}\lambda^{\prime}_{312}$ &~~~~~~~~~~~~~~ $3.55\times 10^{-4}$ &
~~~~~~~~~~~~~~~~$  $ \\
$\lambda^{\prime}_{213}\lambda^{\prime}_{313}$ &~~~~~~~~~~~~~~ $3.60\times 10^{-4}$ &
~~~~~~~~~~~~~~~~$  $ \\
$\lambda^{\prime}_{221}\lambda^{\prime}_{321}$,~~$\lambda^{\prime}_{222}\lambda^{\prime}_{322}$ &~~~~~~~~~~~~~~ $3.56\times 10^{-4}$ &
~~~~~~~~~~~~~~~~$  $ \\
$\lambda^{\prime}_{223}\lambda^{\prime}_{323}$ &~~~~~~~~~~~~~~ $3.61\times 10^{-4}$ &
~~~~~~~~~~~~~~~~$  $ \\
$\lambda^{\prime}_{231}\lambda^{\prime}_{331}$,~~$\lambda^{\prime}_{232}\lambda^{\prime}_{332}$ &~~~~~~~~~~~~~~ $5.80\times 10^{-3}$ &
~~~~~~~~~~~~~~~~$ $ \\
$\lambda^{\prime}_{233}\lambda^{\prime}_{333}$ &~~~~~~~~~~~~~~ $7.48\times 10^{-3}$&
~~~~~~~~~~~~~~~~$  $ \\
\hline
 \end{tabular}\label{dlambda}
 \end{table}

\begin{figure}[tb]
\scalebox{0.9}{\epsfig{file=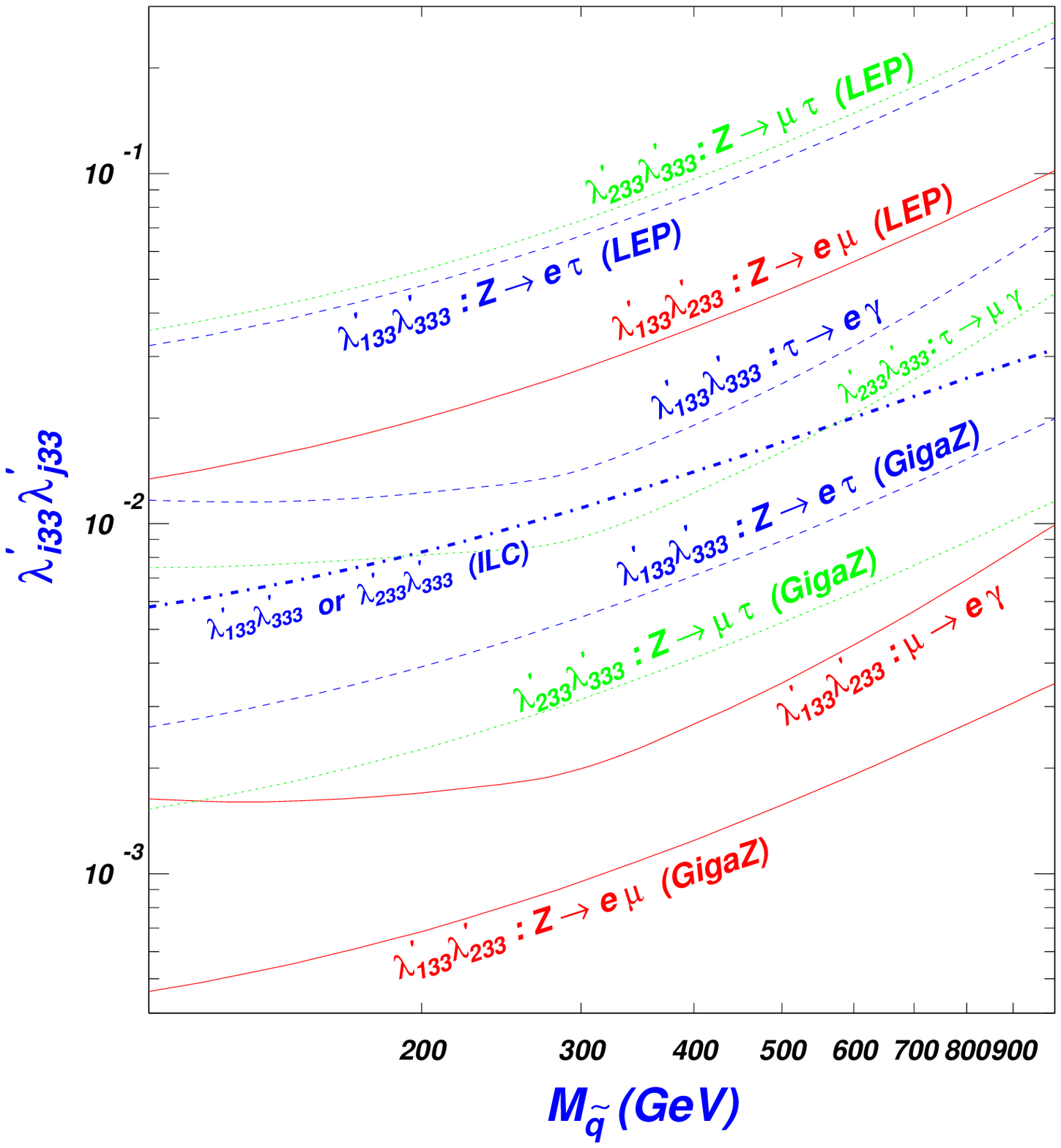}} \vspace*{-0.7cm}
\caption{Various bounds on the $L$-violating couplings versus the
squark mass. The solid, dashed and dotted curves are the bounds on
$\lambda^{\prime}_{133}\lambda^{\prime}_{233}$,
$\lambda^{\prime}_{133}\lambda^{\prime}_{333}$ and
$\lambda^{\prime}_{233}\lambda^{\prime}_{333}$, respectively. Also
shown are  the $2\sigma$ sensitivity from
        $Z$-decays at GigaZ and the 3$\sigma$ sensitivity from
        $\gamma \gamma \to e ({\rm or}~ \mu) ~ \tau$ at the ILC
with center-of-mass
        energy of 500 GeV and a luminosity of $3.45\times 10^{2} fb^{-1}$.}
\label{fig3}
\end{figure}
Note that the neutrino masses could also constrain the
$\lambda^\prime$ couplings, especially $\lambda^\prime_{i33}$
\cite{neutrino-mass-constr}. But these constraints depend on more
parameters in addition to the squark mass. For example, the
one-loop $\lambda^\prime$ contributions to the neutrino masses are
sensitive to the left-right squark mixings and the two-loop
contributions further involve the slepton mass. For small squark
mixings with appropriate sign, there may exist a strong
cancellation between one-loop and two-loop effects, and in this
case, the constraints from the neutrino masses can be avoided.
Since the aim of our study is the sensitivity of the LFC processes
to $\lambda^\prime$ couplings and the $\lambda^\prime$
contributions to these LFC processes are irrelevant to the
additional parameters involved in the contributions to
 the neutrino masses,  in our analysis we did not consider such
constraints from the neutrino masses.

For $Z \to \ell_i \bar\ell_j$, the upper limits from LEP are
\cite{LEP1,LEP2}
\begin{eqnarray}
BR(Z \to \mu e)&<&1.7\times 10^{-6},\\
BR(Z \to \tau e)&<&9.8\times 10^{-6},\\
BR(Z \to \tau \mu)&<& 1.2\times 10^{-5}.
\end{eqnarray}
The bounds from these LEP data are compared with the bounds from
$\ell_i \to \ell_j \gamma$ in Fig.3. One can see that the upper
bounds on the couplings from the LEP $Z$-decay data \cite{LEP1,LEP2}
are weaker than the ones from $\ell_i \to \ell_j \gamma$ data
\cite{exp4}. Note that the bounds from the LEP $Z$-decay data were
also studied in \cite{Z-decay} and our results are consistent with
theirs except that in \cite{Z-decay} the sum over index $k$ is
implied for $\lambda^{\prime}_{i3k}\lambda^{\prime}_{j3k}$ with
$m_{\tilde{q}}=200$ GeV.

The possible sensitivity of GigaZ to the LFC decays of $Z$-boson could reach \cite{gigaz}
\begin{eqnarray}
BR(Z \to \mu e)&\sim &2.0\times10^{-9},\\
BR(Z \to \tau e)&\sim &\kappa\times6.5\times10^{-8},\\
BR(Z \to \tau \mu)&\sim & \kappa\times2.2\times10^{-8}
\end{eqnarray}
with the factor $\kappa$ ranging from 0.2 to 1.0. In Fig. 3 we
take $\kappa=1.0$ to show the sensitivity. In contrast to the
$R$-conserving case in which only $Z \to \mu \tau$ is accessible
at the GigaZ \cite{LFC-1}, the $R$-violating couplings under the
bound from $l_i \to l_j \gamma$ can still enhance all the channels
$Z \to \ell_i \overline{\ell}_j$ to the sensitivity of the GigaZ.
This implies that the GigaZ can further strengthen the bounds on
$\lambda^\prime_{i33} \lambda^\prime_{j33} $ in case of
un-observation. These bounds, unlike the constraints from neutrino
masses which involve more parameters, are only dependent on the
squark mass.

For the $\gamma \gamma$ collision results shown in Fig. 3, we fixed
the parameters as $\xi=4.8$, $D(\xi)=1.83$ and $x_{max}=0.83$
\cite{photon collider}.  Since the $L$-violating couplings relevant
to the process $\gamma \gamma \to e \bar\mu$ is stringently
constrained by $\mu \to e \gamma$,  we in Fig. 3 only show the
results for the channels with a tau lepton in the final states,
i.e., $\gamma \gamma \to e \bar\tau, ~\mu \bar\tau$. The background
for $\gamma \gamma \to e \bar{\tau}$ comes from $\gamma\gamma \to
\tau^{+}\tau^{-} \to \tau^{-}\nu_{e}\bar{\nu}_{\tau}e^{+}$,
$\gamma\gamma \to W^{+}W^{-} \to
\tau^{-}\nu_{e}\bar{\nu}_{\tau}e^{+}$ and $\gamma\gamma \to
e^{+}e^{-}\tau^{+}\tau^{-}$, and we make kinematical cuts 
\cite{l-prod-ILC}: $|\cos\theta_\ell|<0.9$ and $p^{\ell}_{T}>20{\rm
~GeV}$ ($\ell=e,\mu$), to enhance the ratio of signal to background. 
With these cuts, the background cross sections from
$\gamma\gamma \to \tau^{+}\tau^{-} \to
\tau^{-}\nu_{e}\bar{\nu}_{\tau}e^{+}$, $\gamma\gamma \to W^{+}W^{-}
\to \tau^{-}\nu_{e}\bar{\nu}_{\tau}e^{+}$ and $\gamma\gamma \to
e^{+}e^{-}\tau^{+}\tau^{-}$ at $\sqrt{s}=500$ GeV are suppressed
respectively to $9.7\times 10^{-4}$ fb, $1.0\times 10^{-1}$ fb and
$2.4\times 10^{-2}$ fb (see Table I of \cite{l-prod-ILC}). To get the
$3 \sigma$ observing sensitivity with $3.45 \times 10^2$ fb$^{-1}$
integrated luminosity \cite{tesla}, the production rates of $\gamma
\gamma \to e\bar{\tau}, \mu \bar{\tau}$ after the cuts must be
larger than $2.5\times 10^{-2}$ fb \cite{l-prod-ILC}.
We see from Fig. 3 that under the
current bounds from $l_i \to l_j \gamma$, the 
$L$-violating couplings can still be large enough to 
enhance the productions $\gamma\gamma \to e\bar{\tau}, \mu \bar{\tau}$
to the $3 \sigma$ sensitivity.

We also show the cross sections of $\gamma \gamma \to \ell_i
\bar\ell_j$ as a function of center-of-mass energy $\sqrt{s}$ of
the ILC in Fig.4. We see that with the increasing of the
center-of-mass energy, the cross sections of these processes
become smaller. Such a behavior is similar to the results in the
R-conserving MSSM shown in \cite{l-prod-ILC}.

\begin{figure}[tb]
\scalebox{0.6}{\epsfig{file=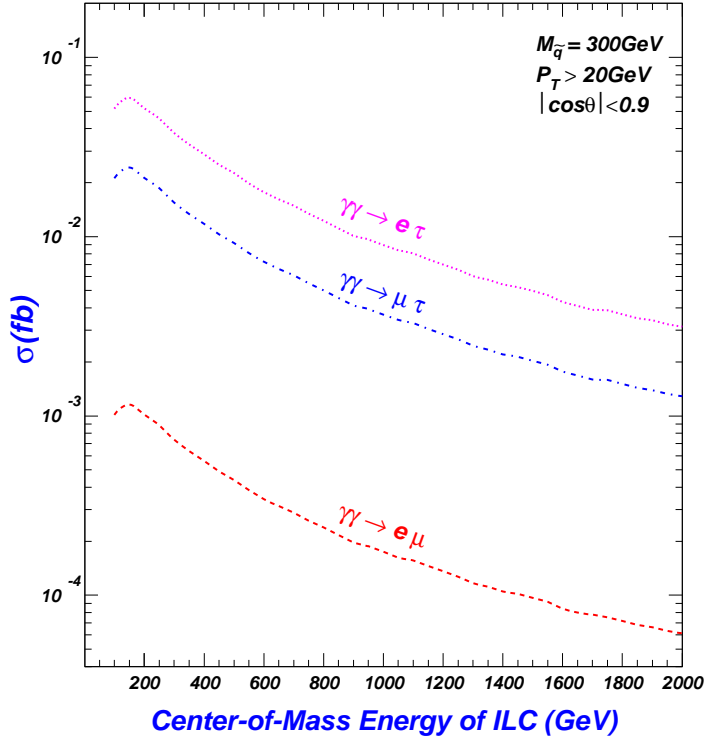}}
\vspace*{-0.4cm}
\caption{The cross sections of $\gamma \gamma \to \ell_i \bar\ell_j$ as a function of
center-of-mass energy $\sqrt{s}$. The couplings $\lambda^{\prime}_{133}\lambda^{\prime}_{233}$,
$\lambda^{\prime}_{133}\lambda^{\prime}_{333}$ and
$\lambda^{\prime}_{233}\lambda^{\prime}_{333}$ are fixed at their upper bounds
at $M_{\tilde{q}}=300{\rm ~GeV}$.}
\label{fig4}
\end{figure}

Finally, we point out that the LFC processes can also put
bounds on the products $\lambda^\prime_{i31} \lambda^\prime_{j31}$
and $\lambda^\prime_{i32} \lambda^\prime_{j32}$, and our numerical
results indicate that such bounds are quite similar to those in
Fig.3. We note that these bounds on  $\lambda^\prime_{i31}
\lambda^\prime_{j31}$ and $\lambda^\prime_{i32}
\lambda^\prime_{j32}$ from $Z \to l_i \bar l_j$ at GigaZ are generally
stronger than those from the neutrino masses \cite{neutrino-mass-constr}.

\section{CONCLUSION}
We evaluated the lepton flavor-changing processes in $R$-parity
violating MSSM. First, we used the latest data on the rare decays
$\ell_i\to \ell_j \gamma$ to update the constraints on the
relevant $R$-violating couplings. Then we calculated the processes
$Z \to \ell_i \bar\ell_j$ and $\gamma\gamma\to \ell_i\bar\ell_j$.
We found that with the updated  constraints the  $R$-violating
couplings can still enhance the rates of these processes to the
sensitivity of GigaZ and photon-photon collision options of the
ILC. So, the GigaZ and photon-photon collision of the ILC can
either observe these $\lambda^\prime$-induced LFC processes or
further strengthen the bounds on the $\lambda^\prime$ couplings in
case of un-observation.

\section*{Acknowledgement}
This work was supported in part by the National Natural Science
Foundation of China (NNSFC) under grant Nos. 10505007, 10821504,
10725526 and 10635030, and by HASTIT under grant No. 2009HASTIT004.

\appendix
\section{Expressions of effective vertex $\gamma(Z)-\ell_i-\ell_j$}
Here we list the expression for the $L$-violating contributions to the effective vertex
$\gamma(Z)-e-\mu$. Other effective vertices $\gamma(Z)-\mu-\tau$ and $\gamma(Z)-e-\tau$
are similar to $\gamma(Z)-e-\mu$ and can be obtained by replacing the corresponding
momentum and mass.
The the effective vertex $\gamma(Z)-e-\mu$ is given by
\begin{eqnarray}
\Gamma_{\lambda}^{\gamma(Z) e \mu}&=& \Gamma_{\lambda}^{\gamma(Z) e \mu}(\tilde {u}_j^L)+
                      \Gamma_{\lambda}^{\gamma(Z) e \mu}(\tilde {d}_k^R) ,
\end{eqnarray}
where the two terms on the right side
denote the $L$-violating loop contributions
by exchanging respectively the squarks $\tilde {u}_j^L$ and $\tilde {d}_k^R$,
given by
\small
\begin{eqnarray}
\Gamma_{\lambda}^{\gamma e \mu}(p_{\mu},p_{e})|_{\tilde {u}_j^L}&=& ae\biggl\{-\frac{1}{3}
[C^1_{\alpha\beta}\gamma^\alpha\gamma_\lambda\gamma^\beta
  -C^1_\alpha (\pslash_{\mu}-\pslash_{e})\gamma_{\lambda}
  \gamma^\alpha ]P_L
 +\frac{2}{3}[2C^2_{\alpha\lambda}\gamma^\alpha
    -C^2_{\alpha}(p_{\mu}+p_{e})_\lambda\gamma^\alpha ]P_L\nonumber\\
&& -\frac{1}{m_{\mu}^2}\gamma^\lambda
     \pslash_{\mu}\gamma^\alpha B^1_{\alpha}P_L
 +\frac{1}{m_{\mu}^2}[\gamma^\alpha \pslash_{e} \gamma_\lambda P_L
      + m_{\mu}\gamma^\alpha \gamma_\lambda P_R)]B^2_{\alpha}
 -\frac{1}{3} m^{2}_{d_{k}}C^1_{0}\gamma_\lambda P_L  \biggl\}\\
\Gamma_{\lambda}^{\gamma e \mu}(p_{\mu},p_{e})|_{\tilde {d}_k^R}&=& ae\biggl\{-\frac{2}{3}
  [C^3_{\alpha\beta}\gamma^\alpha\gamma_\lambda\gamma^\beta
  -C^3_\alpha (\pslash_{\mu} -\pslash_{e})
  \gamma_{\lambda} \gamma^\alpha ]P_L
  +\frac{1}{3}[2C^4_{\alpha\lambda}\gamma^\alpha
    -C^4_{\alpha}(p_{\mu}+p_{e})_\lambda\gamma^\alpha ]P_L\nonumber\\
&&  -\frac{1}{m_\mu^2}\gamma^\lambda
     \pslash_{\mu}\gamma^\alpha B^3_{\alpha}P_L
   +\frac{1}{m_{\mu}^2}[\gamma^\alpha \pslash_{e} \gamma_\lambda P_L
   +m_{\mu}\gamma^\alpha \gamma_\lambda P_R)]B^4_{\alpha}
   -\frac{2}{3}m^{2}_{u_{j}}C_{0}^{3}\gamma_{\lambda} P_{L} \biggl\} \\
\Gamma_{\lambda}^{Z e \mu}(p_{\mu},p_{e})|_{\tilde {u}_j^L}&=& be\biggl\{\frac{2s_w^2}{3}
   [C^5_{\alpha\beta}\gamma^\alpha\gamma_\lambda\gamma^\beta
   -C^5_\alpha (\pslash_{\mu} -\pslash_{e})
   \gamma_{\lambda} \gamma^\alpha ]P_L-(1-\frac{2s_w^2}{3})
   m_{d_k}^2C^5_0\gamma_\lambda P_L\nonumber\\
&& +(1-\frac{4s_w^2}{3})[2C^6_{\alpha\lambda}\gamma^\alpha
   -C^6_{\alpha}(p_{\mu}+p_{e})_\lambda\gamma^\alpha ]P_L
   -\frac{1-2s_w^2}{m_{\mu}^2}\gamma^\lambda
   \pslash_{\mu} \gamma^\alpha B^5_{\alpha}P_L\nonumber\\
&& +\frac{1}{m_{\mu}^2}[(1-2s_w^2)\gamma^\alpha \pslash_{e}\gamma_\lambda P_L
   - 2s_w^2m_{\mu}\gamma^\alpha \gamma_\lambda P_R)]B^6_{\alpha}\biggl\}\\
\Gamma_{\lambda}^{Z e \mu}(p_{\mu},p_{e})|_{\tilde {d}_k^R}&=& be\biggl\{-(1-\frac{4s_w^2}{3})
   [C^7_{\alpha\beta}\gamma^\alpha\gamma_\lambda\gamma^\beta
  -C^7_\alpha (\pslash_{\mu} -\pslash_{e} )
  \gamma_{\lambda} \gamma^\alpha ]P_L+\frac{4s_w^2}{3}m_{u_j}^2C^7_0
  \gamma_\lambda P_L\nonumber\\
&& -\frac{2s_w^2}{3}[2C^8_{\alpha\lambda}\gamma^\alpha
    -C^8_{\alpha}(p_{\mu}+p_{e})_\lambda\gamma^\alpha ]P_L
    -\frac{1-2s_w^2}{m_{\mu}^2}\gamma^\lambda
    \pslash_{\mu} \gamma^\alpha B^7_{\alpha}P_L\nonumber\\
&& +\frac{1}{m_{\mu}^2}[(1-2s_w^2)\gamma^\alpha \pslash_{e} \gamma_\lambda P_L
      - 2s_w^2m_{\mu}\gamma^\alpha \gamma_\lambda P_R)]B^8_{\alpha} \biggl\}
\end{eqnarray}
\normalsize
with $a=\frac{i3\lambda^{\prime}_{1jk}\lambda^{\prime}_{2jk}}{16\pi^2}$,
$b=\frac{i3\lambda^{\prime}_{1jk}\lambda^{\prime}_{2jk}}{16\pi^22s_w c_w}$
and $p_{e}$ and $p_{\mu}$ denoting respectively the momenta of the electron and muon.
In the above expressions, the functions $B^{i}_{\alpha}$ and $C^{i}_{\alpha,\alpha\beta}$
are the Passarino-Veltman functions. For these loop functions, we adopt
the definition in \cite{loop} and use LoopTools \cite{Hahn} in the
calculations. The functional dependence of these  loop functions is given by
\begin{eqnarray}
&& C^1(-p_{\mu},p_{e},m^2_{d_k},m^2_{\tilde{u}_j^L},m^2_{d_k}),
   ~C^2(-p_{e},p_e-p_{\mu},m^2_{d_k},m^2_{\tilde{u}_j^L},m^2_{\tilde{u}_j^L}),\\
&& C^3(-p_{\mu},p_{e},m^2_{u_j},m^2_{\tilde{d}_k^R},m^2_{u_j}),
   ~C^4(-p_{e},p_{e}-p_{\mu},m^2_{u_j},m^2_{\tilde{d}_k^R},m^2_{\tilde{d}_k^R})~,\\
&& C^5(-p_{\mu},p_{e},m^2_{d_k},m^2_{\tilde{u}_j^L},m^2_{d_k}),
   ~C^6(-p_{e},p_e-p_{\mu},m^2_{d_k},m^2_{\tilde{u}_j^L},m^2_{\tilde{u}_j^L}),\\
&& C^7(-p_{\mu},p_{e},m^2_{u_j},m^2_{\tilde{d}_k^R},m^2_{u_j}),
   ~C^8(-p_{e},p_{e}-p_{\mu},m^2_{u_j},m^2_{\tilde{d}_k^R},m^2_{\tilde{d}_k^R})~,\\
&& B^1(-p_{\mu},m^2_{d_k}, m^2_{\tilde{u}_j^L}),
   ~B^2(-p_{e},m^2_{d_k}, m^2_{\tilde{u}_j^L})~,\\
&& B^3(-p_{\mu},m^2_{u_j}, m^2_{\tilde{d}_k^R}),
   ~B^4(-p_{e},m^2_{u_j}, m^2_{\tilde{d}_k^R})~,\\
&& B^5(-p_{\mu},m^2_{d_k}, m^2_{\tilde{u}_j^L}),
   ~B^6(-p_{e},m^2_{d_k}, m^2_{\tilde{u}_j^L})~,\\
&& B^7(-p_{\mu},m^2_{u_j}, m^2_{\tilde{d}_k^R}),
   ~B^8(-p_{e},m^2_{u_j}, m^2_{\tilde{d}_k^R})~.
\end{eqnarray}

\section{Expressions of amplitudes for $\gamma \gamma \to \ell_i \bar{\ell_j}$}
The amplitudes of the diagrams in Fig.2(a-i)
are given by
\small
\begin{eqnarray}
\label{B1}
M_{(a)}|_{\tilde {u}_j^L,{\tilde{d}_k^R}}&=& \overline{u}(e)(ie\gamma_{\lambda})\frac{i}{(\pslash_{2}-\pslash_{\mu})}\Gamma_{\rho}^{\gamma e \mu}(-p_{\mu},p_{2}-p_{\mu})|_{\tilde {u}_j^L,{\tilde{d}_k^R}}v(\mu)\epsilon^{\lambda}_{1}\epsilon^{\rho}_{2}\\
\label{B2}
M_{(b)}|_{\tilde {u}_j^L,{\tilde{d}_k^R}}&=& \overline{u}(e)\Gamma_{\lambda}^{\gamma e \mu}(p_{2}-p_{\mu},p_e)|_{\tilde {u}_j^L,{\tilde{d}_k^R}}\frac{i}{(\pslash_{2}-\pslash_{\mu})}(ie\gamma_{\rho})v(\mu)\epsilon^{\lambda}_{1}\epsilon^{\rho}_{2}\\
M_{(c)}|_{\tilde {u}_j^L}&=& -\frac{i}{16\pi^{2}}(\frac{8}{9}e^{2})\lambda^{'}_{1jk}\lambda^{'}_{3jk}\overline{u}(e)C^{9}_{\alpha}\gamma^{\alpha}P_{L}v(\mu)\epsilon_{1}\cdotp\epsilon_{2}\\
M_{(c)}|_{\tilde{d}_k^R}&=& \frac{i}{16\pi^{2}}(\frac{2}{9}e^{2})\lambda^{'}_{1jk}\lambda^{'}_{3jk}\overline{u}(e)C^{10}_{\alpha}\gamma^{\alpha}P_{L}v(\mu)\epsilon_{1}\cdotp\epsilon_{2}\\
M_{(d)}|_{\tilde {u}_j^L}&=&\frac{i}{16\pi^{2}}(\frac{1}{9}e^{2})\lambda^{'}_{1jk}\lambda^{'}_{3jk}\overline{u}(e)\biggl\{  D^{1}_{\alpha\beta\delta}\gamma^{\alpha}\gamma_{\rho}\gamma^{\beta}\gamma_{\lambda}\gamma^{\delta}+D^{1}_{\alpha\beta}\gamma^{\alpha}\gamma_{\rho}\pslash_{2}\gamma_{\lambda}\gamma^{\beta}\nonumber\\&&+D^{1}_{\alpha\beta}(\pslash_{1}+\pslash_{2})\gamma_{\rho}\gamma^{\alpha}\gamma_{\lambda}\gamma^{\beta}+D^{1}_{\alpha}(\pslash_{1}+\pslash_{2})\gamma_{\rho}\pslash_{2}\gamma_{\lambda}\gamma^{\alpha}+m^{2}_{d_k}D^{1}_{\alpha}(\gamma^{\alpha}\gamma_{\rho}\gamma_{\lambda} \nonumber\\
&&+\gamma_{\rho}\gamma_{\lambda}\gamma^{\alpha}+\gamma_{\rho}\gamma^{\alpha}\gamma_{\lambda})+m^{2}_{d_k}D^{1}_{0}[(\pslash_{1}+\pslash_{2})\gamma_{\rho}\gamma_{\lambda}+\gamma_{\rho}\pslash_{2}\gamma_{\lambda}] \biggl\}P_{L}v(\mu)\epsilon^{\lambda}_{1}\epsilon^{\rho}_{2}, \\
M_{(e)}|_{\tilde{u}_j^L}&=&\frac{i}{16\pi^{2}}(\frac{4}{9}e^{2})\lambda^{'}_{1jk}\lambda^{'}_{3jk}\overline{u}(e)\biggl\{4D^{2}_{\rho\lambda\alpha}-2D^{2}_{\rho\alpha}(2p_e-2p_1-p_2)_{\lambda}\nonumber\\&&-2D^{2}_{\lambda\alpha}(2p_e-p_1)_{\rho}+D^{2}_{\alpha}(2p_e-p_1)_{\rho}(2p_e-2p_1-p_2) \biggl\}\gamma^{\alpha}P_{L}v(\mu)\epsilon^{\lambda}_{1}\epsilon^{\rho}_{2} \\
M_{(f)}|_{\tilde{u}_j^L}&=&\frac{i}{16\pi^{2}}(-\frac{2}{9}e^{2})\lambda^{'}_{1jk}\lambda^{'}_{3jk}\overline{u}(e)\biggl\{[2D^{3}_{\alpha\beta\lambda}-D^{3}_{\alpha\beta}(2p_e-p_2)_{\lambda}]\gamma^{\alpha}\gamma_{\rho}\gamma^{\beta}-[2D^{3}_{\alpha\lambda}\nonumber\\&&-D^{3}_{\alpha}(2p_e-p_2)_{\lambda}]\gamma^{\alpha}\gamma_{\rho}\pslash_{1}+m^{2}_{d_k}[2D^{3}_{\lambda}-D^{3}_0(2p_e-p_2)_{\lambda}]\gamma_{\rho} \biggl\}P_{L}v(\mu)\epsilon^{\lambda}_{1}\epsilon^{\rho}_{2}
\end{eqnarray}
\begin{eqnarray}
M_{(g)}|_{\tilde{d}_k^R}&=&\frac{i}{16\pi^{2}}(\frac{4}{9}e^{2})\lambda^{'}_{1jk}\lambda^{'}_{3jk}\overline{u}(e)\biggl\{D^{4}_{\alpha\beta\delta}\gamma^{\alpha}\gamma_{\rho}\gamma^{\beta}\gamma_{\lambda}\gamma^{\delta}+D^{4}_{\alpha\beta}\gamma^{\alpha}\gamma_{\rho}\pslash_{2}\gamma_{\lambda}\gamma^{\beta}\nonumber\\&&+D^{4}_{\alpha\beta}(\pslash_{1}+\pslash_{2})\gamma_{\rho}\gamma^{\alpha}\gamma_{\lambda}\gamma^{\beta}+D^{4}_{\alpha}(\pslash_{1}+\pslash_{2})\gamma_{\rho}\pslash_{2}\gamma_{\lambda}\gamma^{\alpha}+m^{2}_{d_k}D^{4}_{\alpha}(\gamma^{\alpha}\gamma_{\rho}\gamma_{\lambda}\nonumber\\&&+\gamma_{\rho}\gamma_{\lambda}\gamma^{\alpha}+\gamma_{\rho}\gamma^{\alpha}\gamma_{\lambda})+m^{2}_{d_k}D^{4}_{0}[(\pslash_{1}+\pslash_{2})\gamma_{\rho}\gamma_{\lambda}+\gamma_{\rho}\pslash_{2}\gamma_{\lambda}] \biggl\}P_{L}v(\mu)\epsilon^{\lambda}_{1}\epsilon^{\rho}_{2}\\
M_{(h)}|_{\tilde{d}_k^R}&=&\frac{i}{16\pi^{2}}(-\frac{1}{9}e^{2})\lambda^{'}_{1jk}\lambda^{'}_{3jk}\overline{u}(e)\biggl\{4D^{5}_{\rho\lambda\alpha}-2D^{5}_{\rho\alpha}(2p_e-2p_1-p_2)_{\lambda}\nonumber\\&&-2D^{5}_{\lambda\alpha}(2p_e-p_1)_{\rho}+D^{5}_{\alpha}(2p_e-p_1)_{\rho}(2p_e-2p_1-p_2) \biggl\}\gamma^{\alpha}P_{L}v(\mu)\epsilon^{\lambda}_{1}\epsilon^{\rho}_{2}\\
M_{(i)}|_{\tilde{d}_k^R}&=&\frac{i}{16\pi^{2}}(-\frac{2}{9}e^{2})\lambda^{'}_{1jk}\lambda^{'}_{3jk}\overline{u}(e)\biggl\{[2D^{6}_{\alpha\beta\lambda}-D^{6}_{\alpha\beta}(2p_e-p_2)_{\lambda}]\gamma^{\alpha}\gamma_{\rho}\gamma^{\beta}-[2D^{6}_{\alpha\lambda}\nonumber\\
&&-D^{6}_{\alpha}(2p_e-p_2)_{\lambda}]\gamma^{\alpha}\gamma_{\rho}\pslash_{1}+m^{2}_{d_k}[2D^{6}_{\lambda}-D^{6}_0(2p_e-p_2)_{\lambda}]\gamma_{\rho} \biggl\}P_{L}v(\mu)\epsilon^{\lambda}_{1}\epsilon^{\rho}_{2} .
\end{eqnarray}
\normalsize Here the effective vertices appearing in
Eqs.(\ref{B1}) and (\ref{B2}) are defined in Appendix A. The
amplitudes for the diagrams with the two photons exchanged are not
presented here, which can be obtained from the above corresponding
amplitudes with replacement $p_1\leftrightarrow p_2$ and
$\epsilon_1 \leftrightarrow \epsilon_2$. The functional dependence
of the Passarino-Veltman loop functions $C^{i}_{\alpha}$ and 
$D^{i}_{\alpha,\alpha\beta,\alpha\beta\gamma}$ is given by 
\small
\begin{eqnarray}
&& C^9(-p_e,p_\mu+p_e,m^2_{d_k},m^2_{\tilde{u}_j^L},m^2_{\tilde{u}_j^L}),
  ~~~~~  C^{10}(-p_e,p_\mu+p_e,m^2_{u_j},m^2_{\tilde{d}_k^R},m^2_{\tilde{d}_k^R})\\
&& D^1(p_2,p_1,-p_{e},m^2_{d_k},m^2_{d_k},m^2_{d_k},m^2_{\tilde{u}_j^L}),
  ~ D^2(-p_e,p_1,p_2,m^2_{d_k},m^2_{\tilde{u}_j^L},m^2_{\tilde{u}_j^L},m^2_{\tilde{u}_j^L})\\
&& D^3(-p_e,p_2,-p_{\mu},m^2_{d_k},m^2_{\tilde{u}_j^L},m^2_{\tilde{u}_j^L},m^2_{d_k}),
  ~ D^4(p_2,p_1,-p_{e},m^2_{u_j},,m^2_{u_j},m^2_{u_j},m^2_{\tilde{d}_k^R})\\
&& D^5(-p_e,p_1,p_2,m^2_{u_j},m^2_{\tilde{d}_k^R},m^2_{\tilde{d}_k^R},m^2_{\tilde{d}_k^R}),
  ~ D^6(-p_e,p_2,-p_{\mu},m^2_{u_j},m^2_{\tilde{d}_k^R},m^2_{\tilde{d}_k^R},m^2_{u_j})
\end{eqnarray}

\end{document}